\documentstyle[11pt,newpasp,twoside,epsf]{article}
\def\edcomment#1{\iffalse\marginpar{\raggedright\sl#1\/}\else\relax\fi}
\reversemarginpar

\begin{document}
\title{VINCI / VLTI observations of Main Sequence stars}
 \author{Pierrre Kervella}
\affil{ESO, Alonso de Cordova 3107, Casilla 19001, Santiago 19, Chile}
\author{Fr\'ed\'eric Th\'evenin, Pierre Morel, Janine  Provost, Gabrielle  Berthomieu}
\affil{Observatoire de la C\^ote d'Azur, BP 4229, F-06304 Nice, France}
\author{Damien S\'egransan, Didier Queloz}
\affil{Observatoire de Gen\`eve, CH-1290 Sauverny, Switzerland}
\author{Pascal  Bord\'e}
\affil{Observatoire de Paris, 5, pl. Jules Janssen, F-92195 Meudon, France}
\author{Emmanuel Di Folco}
\affil{ESO, Karl-Schwarzschild-Str. 2, D-95748 Garching, Germany}
\author{Thierry Forveille}
\affil{CFHT Corporation, PO Box 1597, Kamuela, HI 96743, USA}
%\affil{The Name of My Institution, The Full Address of My Institution}

\begin{abstract}
Main Sequence (MS) stars are by far the most numerous class in the Universe.
They are often somewhat neglected as they are relatively quiet objects (but
exceptions exist), though they bear testimony of the past and future of our Sun. An
important characteristic of the MS stars, particularly the solar-type ones, is that they
host the large majority of the known extrasolar planets. Moreover, at the bottom of the
MS, the red M dwarfs pave the way to understanding the physics of brown dwarfs and
giant planets.  We have measured very precise angular diameters from
recent VINCI/VLTI interferometric observations of a number of MS stars in the $K$ band,
with spectral types between A1V and M5.5V.
They already cover a wide range of effective temperatures and radii.
Combined with precise {\it Hipparcos} parallaxes, photometry, spectroscopy as well as the
asteroseismic information available for some of these stars, the angular diameters put
strong constraints on the detailed models of these stars, and therefore on the physical
processes at play.
\end{abstract}

\section{Scientific rationale}

The modeling of the MS stars benefits strongly of the additional constraint provided
by the linear photospheric diameter. Thanks to high precision interferometric measurements
with the VLTI (Glindemann et al.~2000), equipped with the VINCI beam combiner
(Kervella et al. 2000; Kervella et al. 2003a), we are in the process of producing
a coherent list of angular diameters of nearby MS stars. The goal of our program is to
refine our knowledge of their basic properties: age, metallicity, helium content, etc...

\section{Interferometric observations}

For all our measurements, we used the VLT Interferometer with its commissioning instrument,
VINCI, a two telescopes beam combiner operating in the K band (2.0-2.2\,$\mu$m).
This instrument measures the squared visibility ($V^2$) of the interferometric fringes. $V^2$ is related to the
angular diameter of the star through the Zernike-Van Cittert theorem. Fig.1 illustrates the $V^2$
measurements that we obtained on Sirius\,A, and the best-fit model that allowed us to derive its
limb darkened angular size: $\theta_{\rm LD} = 6.039 \pm 0.019$\,mas (Kervella et al.~2003c).
Coupled with the Hipparcos parallax of $\pi = 379.22 \pm 1.58$\,mas (Perryman et al.~1997),
this translates into a radius of $R_\star = 1.711 \pm 0.013$\,R$_{\odot}$.
Among the other stars that we have measured (Table 1) figures Proxima (S\'egransan et al. 2003),
that is only slightly larger than Jupiter ($R_\star = 0.145 \pm 0.011$\,R$_{\odot}$).
The linear radii listed in Table 1 were deduced using the {\it Hipparcos} parallaxes.

\begin{table}
\caption{Physical parameters of the stars of our sample.}
\label{table_radii}
\begin{tabular}{lcllll}
\tableline
Name & Spect. & Mass (M$_{\odot}$) & T$_{\rm eff}$ (K) & $\theta_{\rm LD}$ (mas) & R (R$_{\odot}$)\\
\tableline
Proxima & M5.5V & $0.123 \pm 0.006$ & $3006 \pm 100$ & $1.04 \pm 0.08$ & $0.145 \pm 0.011$\\
GJ191 & M1V & $0.281 \pm 0.014$ & $3419 \pm 100$ & $0.69 \pm 0.06$ & $0.291 \pm 0.025$\\
GJ887 & M0.5V & $0.503 \pm 0.025$ & $3645\pm 100$ & $1.39 \pm 0.04$ & $0.491 \pm 0.014$\\
GJ205 & M1.5V & $0.631 \pm 0.031$ & $3894 \pm 100$ & $1.15 \pm 0.11$ & $0.702 \pm 0.063$\\
$\epsilon$\,Eri & K2V & $0.830 \pm 0.100$ & $5100 \pm 100$ & $2.14 \pm 0.03$ & $0.738 \pm 0.010$\\
$\alpha$\,Cen\,B & K1V & $0.907 \pm 0.006$ & $5250 \pm 50$ & $6.00 \pm 0.03$ & $0.863 \pm 0.005$\\
%Sun & G2V & 1.000  & 5770 & & 1.000 \\
$\tau$\,Cet & G8V & $0.820 \pm 0.030$ & $5250 \pm 100$ & $2.05 \pm 0.03$ & $0.804 \pm 0.010$\\
$\alpha$\,Cen\,A & G2V & $1.100 \pm 0.006$ & $5750 \pm 30$ & $8.51 \pm 0.02$ & $1.224 \pm 0.003$\\
Procyon\,A & F5IV-V & $1.420 \pm 0.050$ & $6530 \pm 50$ & $5.45 \pm 0.05$ & $2.048 \pm 0.025$\\
$\beta$\,Pic & A5V & $1.750 \pm 0.100$ & $8200 \pm 200$ & $0.84 \pm 0.06$ & $1.732 \pm 0.123$\\
Fomalhaut & A3V & $2.000 \pm 0.200$ & $8750 \pm 200$ & $2.22 \pm 0.02$ & $1.870 \pm 0.032$\\
Sirius\,A & A1V & $2.070 \pm 0.060$ & $9900 \pm 200$ & $6.04 \pm 0.02$ & $1.711 \pm 0.013$\\
\tableline
\tableline
\end{tabular}
\end{table}

\begin{figure}
%\plotfiddle{sirius_visib.eps}{7cm}{0}{70}{70}{-135}{0}
\plotfiddle{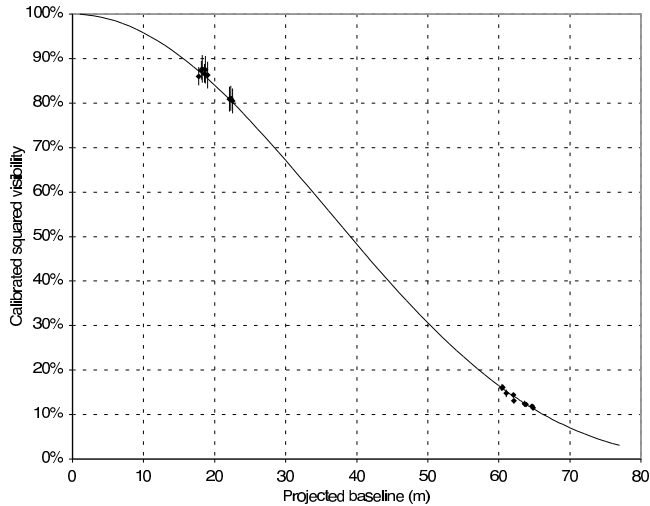}{7cm}{0}{70}{70}{-135}{0}
\caption{Squared visibilities of Sirius A measured with VINCI, and the adjusted
limb-darkened disk model.}
\end{figure}

\section{Modeling using the CESAM code}

To retrieve the properties of these stars (0.9\,M$_{\odot}$ and heavier) we use the
CESAM evolutionary code (Morel 1997), constrained by the widest possible range of observations:
spectroscopy, photometry, astrometry (for binary stars, giving the mass), interferometry,
asteroseismology (when available). For the nearest stars,  combining all these constraints
together yields narrow uncertainties on the model outputs, in particular the age of the stars.
This process benefits strongly of the interferometric constraint provided by the linear diameter.
As shown on Fig. 2, the uncertainty domain defined for Sirius\,A in the HR diagram is reduced
in surface by more than a factor 4, from this constraint alone, as compared to the classical
parameters of temperature and luminosity. On this figure,
the dashed rectangle delimits the uncertainty domain in luminosity and effective
temperature, while the shaded area represents the uncertainty on the interferometric
radius. The continuous line corresponds to a model with
overshoot and the dashed-dot line to a model without overshoot, both with a mass of 
2.12\,M$_{\odot}$ and ages of roughly 200\,Myr. The dashed line corresponds to a
model with a mass of 2.07\,M$_{\odot}$ and an age of 243\,Myr. The modeling of this star is
described in details in Kervella et al.~(2003c).

\begin{figure}
%\plotfiddle{sirius_evol.eps}{6cm}{-90}{35}{35}{-130}{200}
\plotfiddle{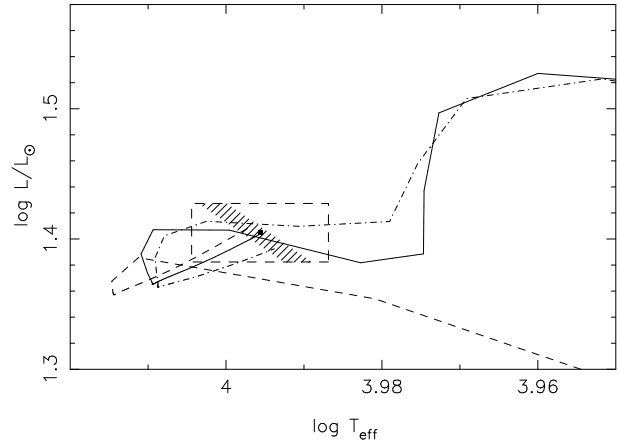}{6cm}{-90}{35}{35}{-130}{200}
\caption{Evolutionary tracks of three models of Sirius\,A in the HR diagram,
computed using the CESAM code (see text and Kervella et al.~2003c for details).
The dashed rectangle delimits the uncertainty domain in
luminosity and effective temperature, while the shaded area represents the
uncertainty on the interferometric radius.}
\end{figure}

\begin{figure}
%\plotfiddle{dwarfs_graph.eps}{8cm}{0}{40}{40}{-170}{0}
\plotfiddle{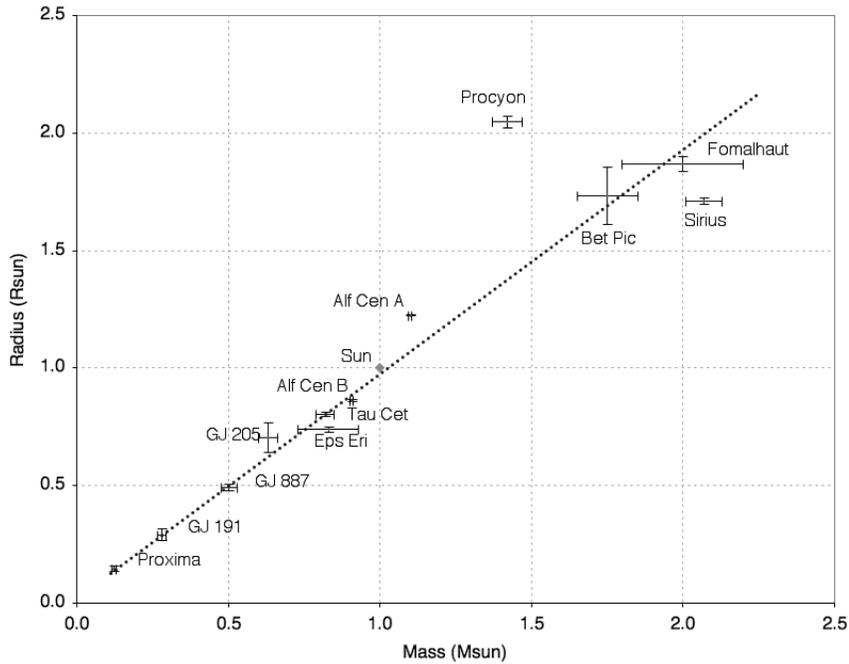}{8cm}{0}{40}{40}{-170}{0}
%\plotone{dwarfs_graph.eps}
\caption{Mass-Radius diagram of the MS stars measured with VINCI/VLTI. The dispersion is due to the
different evolutionary state of each star. $\alpha$\,Cen\,A is already evolved and will soon leave the
Main Sequence.}
\end{figure}

For the very low mass stars (M dwarfs), the evolution is so slow that we currently cannot
deduce precisely their properties by direct modeling of their evolution. However, we have
compared the theoretical Mass-Radius relation predicted for this type of stars to our
measurements (S\'egransan et al. 2003), and the agreement is good even for very low
mass stars ($\le 0.5\,$M$_{\odot}$). This is an important indication that the equation of state
used for these theoretical models is satisfactory, and that the source of the residual
discrepancy lies probably in an imperfect modeling of the energy transport (convection
and opacities).

\section{Perspectives}

One important limiting factor in the determination of linear diameters is the precision
of the {\it Hipparcos} parallax. Though satisfactory in absolute, this precision can be insufficient
compared to the interferometric angular diameter measurement. For $\alpha$\,Cen\,A in
particular, both error sources are of the same order of magnitude, $\pm$0.2\,\%, and the
parallax error contributes significantly to the final uncertainty on the linear size. In this
context, very high precision parallaxes, beyond {\it Hipparcos}, are highly desirable.

Currently, we rely on models of the atmospheric limb darkening to derive the true
photospheric size of the stars, e.g. by Claret (2000). While their precision is relatively
good (approximately $\pm$0.1\,\% of the total diameter), they start to be a limiting factor
in the case of the highest precision diameter measurements. Our objective is to directly
measure the limb darkening of nearby dwarf stars in order to avoid this modeling step. This
implies that we sample the visibility function beyond its first minimum, and therefore
presents particular observational difficulties.
Aditionnally, we are in the process of bridging the gap in terms of spectral types
between $\alpha$\,Cen\,A and Sirius\,A through the observation of nearby F dwarfs.

\end{document}